\documentstyle[aps,preprint,epsfig,floats]{revtex}
\tighten
\begin{document}
\draft

\title{Modeling of protein interaction networks}

\author{Alexei Vazquez$^1$,  Alessandro Flammini$^1$, 
Amos Maritan$^1,2$ and Alessandro Vespignani$^2$}

\address{$^1$INFM-International School for Advanced Studies, via
Beirut 4, 34014 Trieste, Italy\\
$^2$The Abdus Salam International Centre for Theoretical Physics
  (ICTP), P.O. Box 586, 34100 Trieste, Italy }

\date{\today}

\maketitle

\begin{abstract}

We introduce a graph generating model aimed at representing the evolution of
protein interaction networks. The model is based on the hypotesis of
evolution by duplications and divergence of the genes which produce
proteins. The obtained graphs shows multifractal properties recovering the
absence of a characteristic connectivity as found in real data of protein
interaction networks.  The error tolerance of the model to random or
targeted damage is in very good agreement with the behavior obtained in real
protein networks analysis.  The proposed model is a first step in the
identification of the evolutionary dynamics leading to the development of
protein functions and interactions.

\end{abstract}


The complete genome sequencing gives for the first time the means to
analyze organisms on a genomic scale. This implies the understanding of the
role of a huge number of gene products and their interactions.  For
instance, it becomes a fundamental task to assign functions to
uncharacterized proteins, which are traditionally identified on the basis
of their biochemical role. The functional assignment has progressed
considerably by partnering proteins of similar functions, leading
eventually to the drawing of proteins interaction networks (PIN).  This has
been accomplished in the case of the yeast {\em Saccharomices Cerevisiae}
where two hybrid analyses and biochemical protein interaction data have
been used to generate a web-like view of the protein-protein interaction
network\cite{uetz00}.  The topology of the obtained graph has been recently
studied in order to identify and characterize its intricate
architecture\cite{wagner00,jeong01}.  Surprisingly, the {\em S. Cerevisiae}
PIN resulted in a very highly heterogeneous network with scale-free
connectivity properties.

Scale-free (SF) networks opposes to regular (local or non-local) 
networks for the small average 
link distance between any two nodes (small-world property)\cite{watts98}
and a statistically significant probability that nodes possess
a large number of connections $k$ compared to the average connectivity
$\left<k \right>$~\cite{barabasi99}. 
This reflects in a power-law connectivity distribution 
$P(k)\sim k^{-\gamma}$ for the probability that a node has $k$ links 
pointing to other nodes. These topological properties, that appear 
to be realized in many natural and technological 
networks (for a review see Ref.\cite{strog01}), are due to the 
interplay of the network growth and the ``preferential attachment''
rule. In other words, new appearing nodes have a higher probability
to get connected to another node already counting a large number of links.
These ingredients encompassed in the Baraba\'si and Albert  (BA)
model \cite{barabasi99} are present in all the successive 
scale-free network models \cite{barabba00,krap00,mendes00} 
and  are fundamental for the spontaneous evolution of networks in 
a scale-free architecture. 
In this perspective it is natural to ask about the  microscopic 
process that drives the placement of nodes and links in the case of the 
protein network. 

In the present Letter we study a growing network model whose microscopic
mechanisms are inspired by the duplication and the functional
complementation of genes\cite{wagner00,force99}. In this evolutive model
all proteins in a family evolved from a common ancestor through gene
duplications and mutations (divergence), and the protein network is the
blueprint of the entire history of the genome evolution. This
duplication-divergence (DD) model is analyzed by analytical calculations
and numerical simulations in order to characterize the topological and
large scaling properties of the generated networks. We find that the
networks show the absence of any characteristic connectivity and exhibit a
connectivity distribution with multifractal properties. The $n$-th moment
of the distribution behaves as a power law of the network size $N$ with an
exponent that has a nonlinear dependency with $n$. In addition, all moments
crosses from a divergent behavior with the network size to a finite
asymptotic value at a given value of the mutation parameters. The generated
networks can be directly compared with the {\em Saccharomices Cerevisiae} 
PIN data reported in \cite{datasource}, composed by 1,825 proteins (nodes) 
connected by 2238 identified interactions (links). In agreement with
the PIN analysis we find that the connectivity distribution can be
approximately fitted by a single apparent power-law behavior, and the
model's parameter can be optimized in order to reproduce quantitatively
magnitudes such as the average connectivity or the clustering coefficient.  
Further, we compare the stimulated tolerance of the DD network to random
deletion of individual nodes with that obtained with deletion of the most
connected ones. While the DD network proves to be fragile in the latter
case, it results extremely resistent to random damages, in agreement with
the behavior recently observed for the PIN \cite{jeong01}. The analogous
topological properties found in the proposed model and the experimental PIN
represent a first step towards a possible understanding of protein-protein
interactions in terms of genes evolution.

Proteins are divided in families according to their sequence and 
functional similarities \cite{families}. 
The existence of these families can be explained using the evolutive
hypothesis that all proteins in a family evolved 
from a common ancestor\cite{ohno70}. This evolution is thought 
to take place through gene or entire genome 
duplications, gaining redundant genes. After the duplication 
redundant genes diverge evolving to perform different
biological functions, loosing or gaining new interactions 
in the metabolic pathways.
According to the classical model \cite{ohno70} after 
duplication the duplicate genes have fully overlapping 
functions. Later on, one of the copies may either 
become nonfunctional due to degenerative mutations 
or it can acquire a novel beneficial function and become preserved 
by natural selection. In a more recent framework \cite{force99} it
is proposed that both duplicate genes are subject to 
degenerative mutations loosing some
functions but jointly retaining the full set of functions 
present in the ancestral gene.
The outcome of this evolution results in the complex network of 
protein interactions.  
This process can be easily translated in a growing network model.
We can consider each node of the network as the protein expressed 
by a gene. After gene duplication both the expressed proteins  will 
have the same interaction map. This corresponds to the addition of a 
new node in the network with the same links of its ancestor. 
Moreover, if the ancestor is a self-interacting protein the copy will 
have also an interaction with it \cite{wagner00}. 
Eventually, some of the common links will be removed 
because of the divergence process. 
We can formalize this process by defining an evolving network in which 
at each time step a node is added according to the following rules
\begin{itemize}

\item{\em Duplication}: a node $i$ is selected at random. A new node  
$i\prime$ with a link to all the neighbors of $i$ is created. 
With probability $p$ a link between $i$ and $i\prime$ is established 
(self-interacting proteins).

\item{\em Divergence}: for each of the node $j$ linked to 
$i$ and  $i\prime$ we choose one of the two links and remove 
it with probability $q$.

\end{itemize}
For practical purposes the duplication-divergence (DD) algorithm starts 
with two connected nodes and repeat the 
duplication-divergence rules $N$ times.
Since genome evolution analysis \cite{wagner00,huynen98} supports 
the idea that the divergence of duplicate genes takes place shortly after the 
duplication, we can assume that the 
divergence process always occurs before any new duplication takes 
place; i.e.,  we are in presence of a time scale separation between 
duplication and mutation rates. This allows us to consider the 
number of nodes in the network, $N$, 
as a measure of time (in arbitrary units). It is worth remarking that the  
algorithm does not include the creations of new links; i.e. the developing
of new biochemical interactions between gene products.
This process has been argued to have a probability much smaller than 
the divergence one\cite{wagner00}. 
However, we have tested that the introduction in 
the DD algorithm of a probability to develop new random 
connections does not change the network topology. 

\begin{figure}[t]

\centerline{\epsfig{file=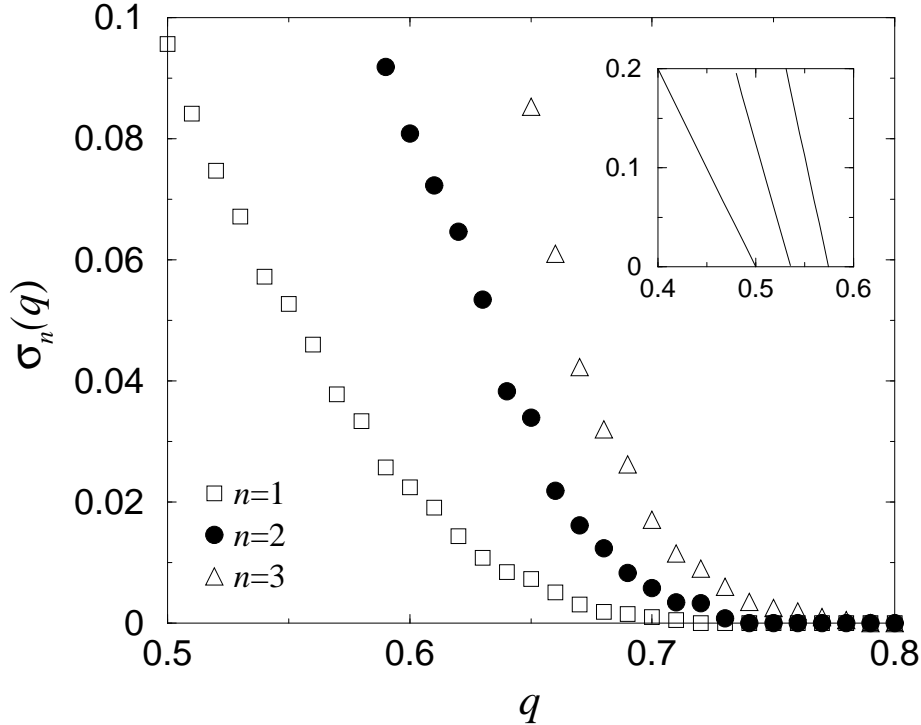,height=10cm}}

\caption{The exponent $\sigma_n(q)$ as a function of $q$ for different 
values of $n$. The symbols were obtained from numerical simulations of 
the model. The moments $\left< k^n\right>$ were computed as a function 
of $N$ in networks with size ranging from $N=10^3$ to $N=10^6$. 
The exponents $\sigma_n(q)$ are obtained from the power 
law fit of the plot $\left< k^n\right>$ vs. $N$. In the inset we 
show the corresponding MF behavior, as obtained from Eq. (\ref{eq:2}), 
which are in qualitative agreement with the numerical results.}

\label{fig:1}

\end{figure}

In order to provide a general analytical understanding of the model, 
we use a mean-field approach for the moments distribution behavior.
Let us define  $\left< k\right>_N$ the average connectivity of the network 
with $N$ nodes.
After a duplication event $N\to N+1$ we have that the average connectivity 
is given by 
\begin{equation}
\left< k\right>_{N+1}=\frac{(N)\left< k\right>_N + 2p+(1-2q)
\left< k\right>_N}{N+1},
\label{eq:1}
\end{equation}
where the r.h.s  represents the average number of links 
gained/lost after the DD process. On average, in fact, there will be a gain 
proportional to $2p$ because of self-interaction wiring and a loss of 
links proportional to $2 q \left< k\right>_{N}$ due to the divergence process. 
For large $N$, taking the continuum limit, we obtain a 
differential equation for $\left<k\right>$. 
For $q>1/2$, $\left< k\right>$ grows with $N$ but saturates to 
the stationary value $k_\infty=2p/(2q-1)+{\cal O}(N^{1-2q})$, 
On the contrary, for $q<1/2$, $\left< k\right>$ grows with $N$ as $N^{1-2q}$. 
At $q=q_1=1/2$ there is a dramatic change of behavior in the large scale 
connectivity properties.  Analogous equations can be written for higher order 
moments $\left< k^n\right>$,
and for all $n$ we find a value $q_n$ at which the  moments crosses
from a divergent behavior to a finite value for $N\to \infty$. 
More interestingly, it is possible to write the generalized exponents
$\sigma_n(q)$ characterizing the moments divergence as 
$\left< k^n\right>\sim N^{\sigma_n(q)}$. The lengthy calculation that will 
be reported elsewhere \cite{next}, gives the mean-field (MF) estimate
\begin{equation}
\sigma_n(q)=n(1-q)-2+2(1-q/2)^n.
\label{eq:2}
\end{equation}
The nonlinear behavior with $n$ is indicative of a multifractal connectivity
distribution.  In order to support the analytical calculations, we performed
numerical simulations generating DD networks with size ranging from $N=10^3$
to $10^6$. In Fig. \ref{fig:1} we report the generalized exponents
$\sigma_n(q)$ a s a function of the divergence parameter $q$. As predicted
by the analytical calculations, $\sigma_n=0$ at a critical value $q_n$.
The general phase diagrams obtained is in good agreement with the MF
predictions and the multifractal picture.

Noticeably, multifractal features are present also in a recently introduced
model of growing networks \cite{doromulti} where, in analogy with the
duplication process, newly added nodes inherit the network connectivity
properties from parent nodes.  Multifractality, thus, appears related to
local inheritance mechanisms. Multifractal distribution have a rich scaling
structure where the scale-free behavior is characterized by a continuum of
exponents. This behavior is, however, opposite to usual exponentially
bounded distribution and it is interesting to understand why in the DD model
we generate the large connectivity fluctuation needed for a scale-free
behavior. A given node with connectivity $k$ receives, with probability
$k/N$, a new link if one of its neighbors is chosen in the duplication
processes. In this case the new added node will establish a new link to it
with probability $(1-q)$. Hence, the probability that the degree of a node
increases by one is
 \begin{equation} 
w_{kN}\sim(1-q)k/N 
\label{eq:3} 
\end{equation}
where we have neglected the constant contribution 
given by the self-interaction probability.
This shows that  even if the evolution rules of the 
DD model are local they introduce an effective linear 
preferential attachment, that is known to be at the origin of scale-free
connectivity distribution \cite{barabasi01,dorogotsev00}. 
However, because the degree and link deletion of duplicate nodes introduce
additional heterogeneity  in the problem we obtain a multifractal 
behavior.

\begin{figure}[t]

\centerline{\epsfig{file=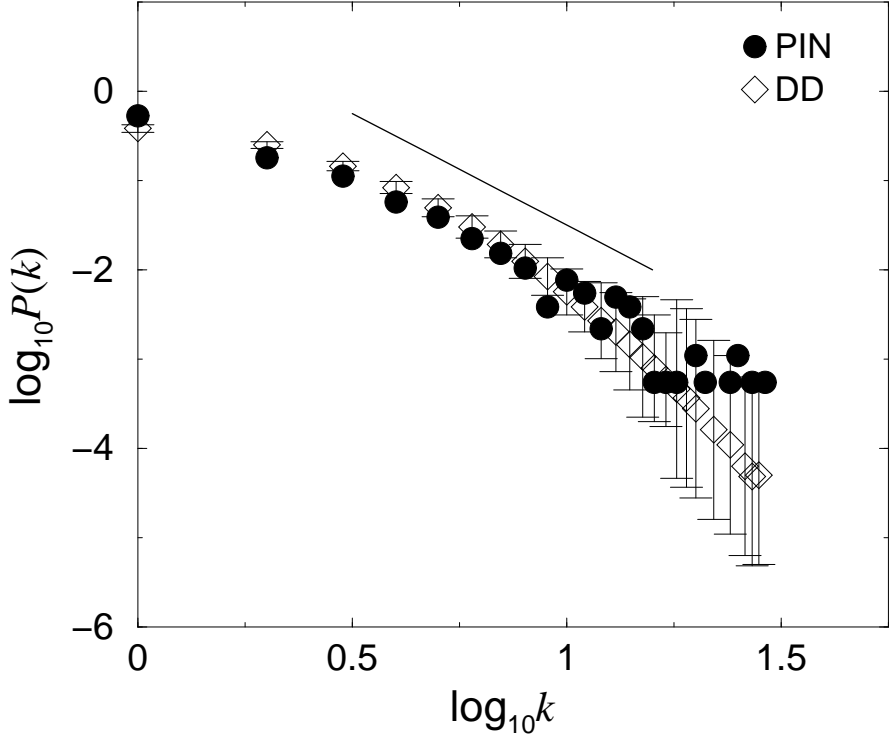,height=10cm}}

\caption{Connectivity distribution for the PIN and the 
DD model with $p=0.1$, $q=0.7$ with $N=1825$. Errors bars represent standard 
deviation on a single network realization.
The straigth line is a power law with exponent $2.5$.}

\label{fig:2}

\end{figure}

The peculiarities of the duplication
and divergence process manifest quantitatively in other features
characterizing the topology of the network. Among these 
the tendency to generate  biconnected triplets and quadruples 
of nodes. These are sets of nodes connected by a simple cycle of links, 
thus forming a triangle or a square.  
In the  DD model triangle formation is a pronounced effect since  
with probability $p(1-q)$ the duplicating genes and any neighbor of the
parent gene will form a new triangle. Analogously, duplicating genes
and any couple of neighbors of the parent gene will form a new square
with probability $(1-q)^2$. An indication of triangles formation
in networks is given by the clustering coefficient 
$C_{\bigtriangleup}=3 N_{\bigtriangleup}/N_{\wedge} $ \cite{newman01}
where $N_{\bigtriangleup}$ is
the number of biconnected triplets (triangles) and $N_{\wedge} $ is the
total number of simply connected triplets.
Similarly it is possible to define the square coefficient
$C_{\Box}=4 N_{\Box}/N_{\Pi}$, with $N_{\Box}$ the number of squares
in the network and $N_{\Pi}$ the number of simply connected quadruples.
By measuring these quantities in the yeast {\em Saccharomices Cerevisiae} 
in \cite{datasource}
PIN, we obtain $C_{\bigtriangleup}=0.23$ and $C_{\Box}=0.11$.
These values are particularly large and difficult to obtain with 
other growing networks model, for which it has been shown that 
the clustering coefficient is algebraically decaying with the network
size. On the contrary, the DD model shows clustering coefficients 
saturating at a finite value, and  it is possible to tune the 
parameters $p$ and $q$ in order to recover the real data estimates. 
A reasonable agreement with the values 
obtained for the real PIN is found when $p\simeq 0.1$ and $q\simeq 0.7$,
which yield networks with $C_{\bigtriangleup}=0.10(5)$ and 
$C_{\Box}=0.10(2)$.
Noticeably, for these values of the parameters the DD model generates 
networks where other quantities, such as the average degree, are in good 
agreement with those obtained from experimental data. A pictorial 
representation of this agreement is provided in Fig. \ref{fig:2}, where  
we compare the connectivity  distribution obtained from  
$10^3$ realizations of the DD model with optimized $p$ and $q$   
and that of  the yeast {\em Saccharomices Cerevisiae} PIN. The DD networks 
are composed by $N=1825$ nodes as for the yeast PIN.  
The agreement is very good, considering the relatively
large statistical fluctuations we have for this network size. 
Error bars on the DD model refers to statistical fluctuations 
on single realizations. It is worth noticing, 
that despite the evident multifractal nature 
of the DD model, for a single realization of  
size consistent with that of the PIN, 
the intermediate $k$ behavior can be approximated by an 
effective algebraic decay with exponent $\gamma\simeq 2.5$ 
as found in Ref.\cite{jeong01}. 

\begin{figure}[t]

\centerline{\epsfig{file=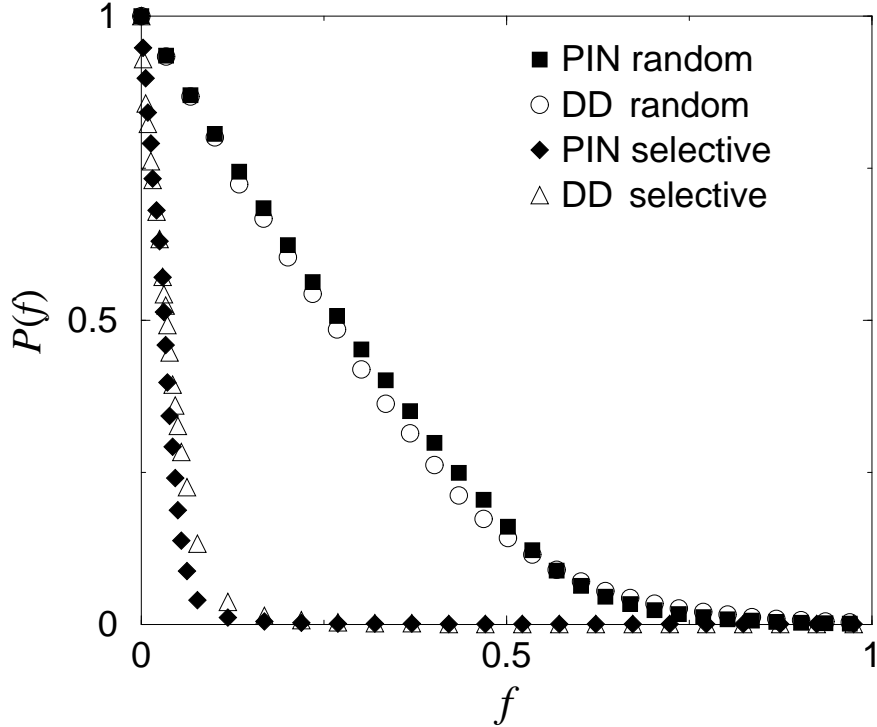,height=10cm}}

\caption{Fraction of nodes $P(f)=N(f)/N$ in the largest network
after a fraction $f$ of the nodes have been removed.
$N(f)$ is defined as the size of the largest network of 
connected sites. Two different removal strategies have been 
used, random and selective (see text). 
In both cases, the DD model curves were obtained after an average 
over $100$ network realizations.}

\label{fig:3}

\end{figure}

Finally, we examine the behavior of the DD model under random and selective 
deletion of nodes and compare it with those obtained for the 
yeast PIN \cite{jeong01}. Resilience to damage is indeed considered an 
extremely relevant property for a network. 
From an applicative point of view it gives a measure of how robust 
a network is against disruptive modifications and how far one can go 
in altering it without destroying its connectivity and therefore 
functionality. 
In the random deletion process of a fraction $f$ of nodes and the  
relative links, it has been observed that the network fragments in 
several disconnected components with a largest connected one of 
size $N(f)$. For random graphs it is known that 
the fraction of nodes  $P(f)=N(f)/N$ belonging to 
the largest remaining network undergoes  an inverse
percolation transition \cite{robust}. In the thermodynamic limit ($N\to 
\infty$), above a fraction $f_c$ of deleted nodes,  
the density $P$ drops to zero; i.e.
no dominant network (giant component) is left.
On the contrary, in scale free networks the density of the largest 
cluster drops to zero only in the limit $f_c\to 1$, denoting an 
high resilience to random damages. Associated with this property, 
it has been observed that scale-free networks are very fragile with 
respect to targeted removal of the highest connected nodes. 
In this case a  small fraction of removed sites fragments completely the 
network ($P\to 0$). A similar behavior has been pointed out to occur also 
in the yeast PIN\cite{jeong01}. 
Fig. \ref{fig:3} shows the density of the largest remaining network versus
the fraction of removed nodes both for PIN and  DD model for random 
and selective nodes removal. The latter case consists in systematically 
removing nodes with the highest degree. 
The DD network tolerance to damage is determined by the scale-free nature 
of its multifractal distribution and the obtained curves are in 
very good agreement with the corresponding ones for the yeast PIN. 
It is worth noticing
again that  the parameters used for the DD network have not been 
independently estimated, but  are those obtained from the 
previous optimization of the clustering coefficients. 
The striking analogies in the tolerance behavior  
are an important test to assess the efficacy of the DD model in 
reproducing the PIN topology.

In conclusion we presented
a physically motivated dynamical model for the network of protein
interactions in biological systems. The model is based on a simple 
process of gene duplication and differentiation, which is believed
to be the main mechanisms beyond the evolution of PIN. Although
the resulting networks share common features with other scale free 
networks, they present novel and intriguing properties, both in 
the degree distribution (multifractality) and in their topology.
The model reproduces with noticeable accuracy the topological
properties of the real PIN of yeast {\em S. Cerevisiae}.  

A. M. and A. F. acknowledge funding from Murst Gfin'99. AV has been
partially supported by the European Network Contract No. ERBFMRXCT980183.

\end{document}